\documentclass[12pt,preprint]{aastex}

\shorttitle{Early Swift Afterglows}
\shortauthors{Butler \& Kocevski}

\def\gtrsim{\mathrel{\hbox{\rlap{\hbox{\lower4pt\hbox{$\sim$}}}\hbox{$>$}}}}
\def\lessim{\mathrel{\hbox{\rlap{\hbox{\lower4pt\hbox{$\sim$}}}\hbox{$<$}}}}
\def\degree{$^{\circ}$}
\newcommand{\beq}{\begin{equation}}
\newcommand{\eeq}{\end{equation}}

\newcommand\swift{{\it Swift}}

\slugcomment{Accepted to ApJ}

\begin{document}

\title{X-ray Hardness Variations as an Internal/External Shock Diagnostic}
\author{Nathaniel R. Butler\altaffilmark{1,2} and
Daniel Kocevski\altaffilmark{2}}
\altaffiltext{1}{Townes Fellow, Space Sciences Laboratory,
University of California, Berkeley, CA, 94720-7450, USA}
\altaffiltext{2}{Astronomy Department, University of California,
445 Campbell Hall, Berkeley, CA 94720-3411, USA}

\begin{abstract}
The early, highly time-variable X-ray emission immediately 
following GRBs exhibits strong spectral variations that are
unlike the temporally smoother emission
which dominates after $t\sim 10^3$ s.  The ratio of hard channel (1.3-10.0 keV)
to soft channel (0.3-1.3 keV) counts in the Swift X-ray telescope
provides a new measure delineating the end time of this emission.
We define $T_{H}$ as the time at which this transition takes place
and measure for 59 events a range of transition times that span $10^2$ s
to $10^{4}$ s,
on average 5 times longer than the prompt $T_{90}$ duration observed in
the Gamma-ray band.
It is very likely that the mysterious light curve plateau phase and  
the later powerlaw temporal evolution, both of which typically occur  
at times greater than $T_{H}$ and hence exhibit very little hardness  
ratio evolution, are both produced by external shocking of the  
surrounding medium and not by the internal shocks thought responsible  
for the earlier emission.
We use the apparent lack of spectral evolution to discriminate  
against proposed models for the plateau phase emission.
We favor energy injection scenarios with a roughly linearly
increasing input energy versus time for six well sampled events with
nearly flat light curves at $t\approx 10^3-10^4$ s.
Also, using the transition time $T_{H}$ as the delineation between the GRB
and afterglow emission, we calculate that the
kinetic energy in the afterglow shock is typically a factor of 10
lower than that released in the GRB.  Three very bright events suggest that
this presents a missing X-ray flux problem rather than an efficiency problem
for the conversion of kinetic energy into the GRB.
Lack of hardness variations in these three events may be due to a very highly relatavistic
outflow or due to a very dense circumburst medium.
There are a handful of rare cases of very late time $t>10^4$ s
hardness evolution, which may point to residual central engine activity
at very late time.
\end{abstract}


\section{Introduction}
\label{sec:intro}

The X-ray telescope \citep{burrows05b} on {\it Swift}~\citep{gehrels04} is opening a
new window into the early lives of $\gamma$-ray Bursts (GRBs) and their afterglows.
Although hints and probable examples of highly time variable X-ray behavior at early
time were seen prior to {\it Swift}~\citep[e.g.,][]{piro05,watson06}, the XRT has
shown us that this behavior is the norm.   Nearly all afterglows show a period
of rapid flux decline after the prompt or flare emission and about half show bright
X-ray flares \citep[e.g.,][]{burrows07}.  How these observations are to be reconciled
with the well-tested internal/external shock GRB and afterglow model \citep{rnm94,snp97,spn98,wng99} ---
which explained very well pre-{\it Swift}~observations of simple fading powerlaws at late time ---
comprises a set of key open questions.

An accurate accounting of the GRB and afterglow phenomenology is critical for comparison to
the models.
\citet{obrien06} have shown that the late GRB as measured by the Burst Alert Telescope
(BAT) transitions in time smoothly into the early X-ray counts detected by the XRT,
provided a correction is made for the different energy bands.  This
demonstrates a close connection between the early X-ray emission and that from the GRB.
In \citet[][BK07]{bNk07}, we fit the BAT and XRT spectra to explicitly show that the best-fit models at early time
are those which fit GRB spectra well.  The early X-ray spectra look like GRB spectra (but have
$\nu F_{\nu}$ peak energies $E_{\rm peak}$ in the X-ray band rather than the $\gamma$-ray band)
and evolve spectrally in a similar fashion \citep[see, also,][]{falcone06,godet07}.
Combined with studies in the time domain indicating
fine timescale variability \citep[e.g.,][]{burrows05a,falcone06,romano06,pagani06,koc07},
we are becoming confident that the X-ray emission prior to about $10^3$s is due to the GRB.
The flat or ``plateau phase'' light curve which is typically present after this phase remains, however,
largely mysterious.

Several models have been proposed to explain the plateau phase light curve.  Because it
is difficult to produce so flat a decay in the external shock picture, the energetics
may be driven by a re-injection from the central engine or late time internal shocks
\citep[e.g.,][]{nousek06,zhang06,gglf07,pain06,pain07}.
Off-axis external shocks \citep{eg06}, the reverse shock \citep{gdm07,ub07}, or
time-varying microphysical parameters \citep{gkp06} may also be responsible.

Observationally, \citet{willin06} have shown that the prompt
and afterglow emission can be separated near the start of the plateau
by fitting two models (of the same form) to each
inferred component.  This split falls short of decisive because
\citet{willin06} are unable to measure the rise of the afterglow component from under the prompt component.
The number of degrees of freedom in the model is large (6--8) and comparable 
to the typical number of powerlaw segments in broken powerlaw fits,
which assume no separation into prompt and afterglow components.
As we show below, a cleaner separation that requires no manual 
removal of flare-like emission
is possible if we consider the spectral variations at early time.  
This can be demonstrated through  
the use of time resolved spectroscopy as discussed above, although  
such efforts are limited to bright events with high signal-to-noise.
Because spectral fits are not required, variations in the X-ray hardness
ratio provides an alternative.  Studying the hardness ratio, we can link the 
plateau phase emission to late-time
external shock emission for even faint bursts.  

After a brief review
of the X-ray phenomenology versus time gleaned from powerlaw fits (Section \ref{sec:temp_spec}), we discuss 
 in Section \ref{sec:endhard} how the hardness evolution implies a separation between prompt and afterglow emission.  
 Stable hardness ratios during and after the plateau phase are exploited to constrain the GRB and afterglow
models in Sections \ref{sec:achromat} and \ref{sec:discussion}.

\section{Data Reduction}

We download the {\it Swift}~XRT data from the 
{\it Swift}~Archive\footnote{ftp://legacy.gsfc.nasa.gov/swift/data}.
The data are processed with version 0.10.3 of the {\tt xrtpipeline} 
reduction script from the 
HEAsoft~6.0.6\footnote{http://heasarc.gsfc.nasa.gov/docs/software/lheasoft/}
software release.  We employ the latest (2006-12-19) calibration 
files.  The reduction of XRT data from cleaned event lists 
output by {\tt xrtpipeline} to science ready light curves and spectra is 
described in detail in Butler \& Kocevski (2007).   Our final
light curves have a fixed signal-to-noise of 3 in the 0.3--10.0 keV band.

We define an X-ray hardness ratio $HR$ as the fraction of counts in
the 1.3--10.0 keV band to the counts in the 0.3--1.3 keV band.  On
average, this ratio is equal to unity for XRT data.  The mean
energy index (flux proportional to $E^{-\beta}$) is $\beta=1$ \citep{butler07a}.
We show in BK07 that the column densities $N_H$, as inferred from soft X-ray
absorption, do not appear to change in time for {\it Swift}~afterglows.
To lowest order for a typical column density $N_H=10^{21}$ cm$^{-2}$,
$\beta \approx 1-0.9\log_e(HR)$.

\subsection{Light Curve Region Selection and Fitting}

To group the data
into separate regions of similar temporal and spectral evolution,
we fit the data using an extension of the Bayesian blocks algorithm
\citep{scargle98} to piecewise logarithmic data.  Our implementation
is simple and requires no human intervention.  Considering each data
point as the location of a possible powerlaw break in the light curve,
we calculate $\chi_{\nu}^2$ for every possible connection of the data points.
This search can be done efficiently using publicly available Bayesian blocks 
code\footnote{http://space.mit.edu/CXC/analysis/SITAR/}.
Because we also include an additional term (a prior against the break)
with each possible new segment of $\Delta \chi_{\nu}^2=9.2$, each new segment must
improve the fit at $\gtrsim 99$\% confidence.  The $\Delta \chi_{\nu}^2$
additions exclude models with many breaks, while the final fit without
the $\Delta \chi_{\nu}^2$ contributions has $\chi_{\nu}^2 \approx \nu$
and typically 3--5 powerlaw segments.

So that noise fluctuations in the data do not generate spurious light
curve breaks, we denoise the soft and hard channel light curve with Haar 
wavelets \citep[see,][]{kol00} prior to fitting.
We fit the count rate and hardness simultaneously so that spectrally
distinct regions are separated.  The fits are plotted in red in
the figures below.  Light curves and fits for all {\it Swift}~events
can be downloaded from\footnote{http://astro.berkeley.edu/$\sim$nat/swift}.

\section{Temporal/Spectral Overview of Swift and pre-Swift X-ray Afterglows}
\label{sec:temp_spec}

The X-ray spectral and temporal properties
of GRB afterglows are typically garnered from powerlaw fits in time and energy.  
For {\it Swift}~data, this can be done
for individual events at multiple epochs.  A detailed study
of the evolution across all of these snapshots is beyond the scope of this paper, and
we restrict our analysis in this section to the time regions as though they were separate events.
It is reasonable to compare {\it Swift}~data to pre-{\it Swift}~data in this fashion,
because pre-{\it Swift} fits correspond to a narrow
time windows in the life of each afterglow and do not typically allow for 
time evolution studies for individual bursts.
We determine fit regions 
with the blocking algorithm described above, and the results are binned
in time into the three time epochs.

\begin{figure}[H]
\centerline{\includegraphics[width=7.0in]{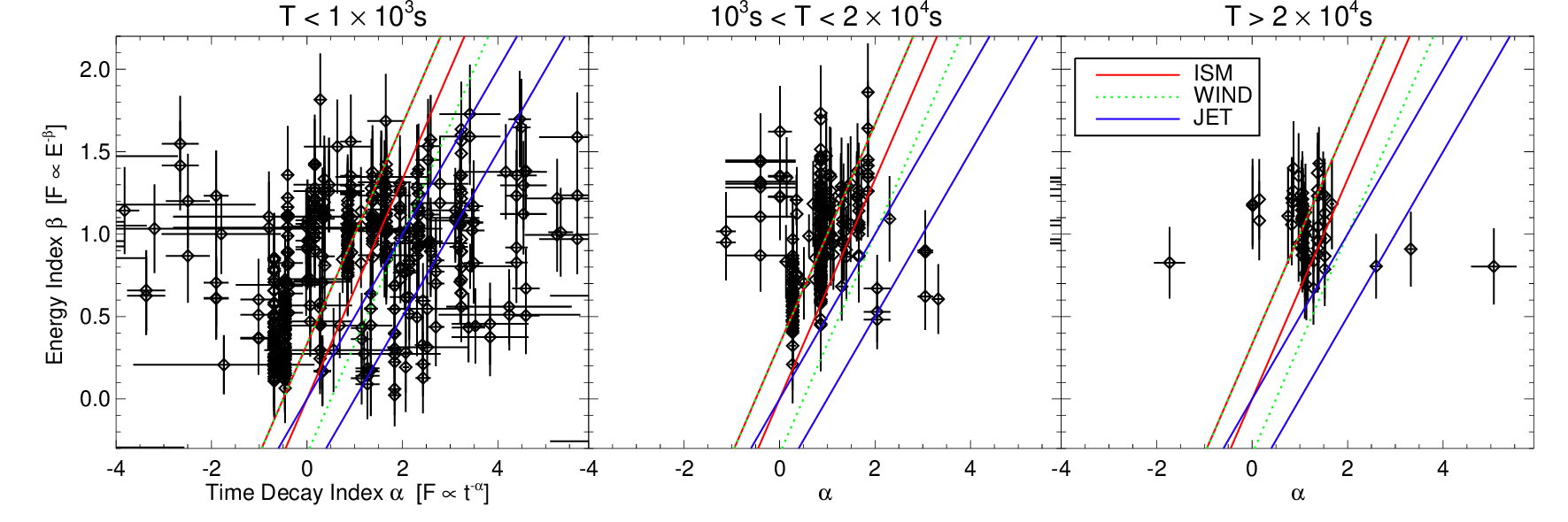}}
\caption{
Powerlaw energy index versus powerlaw light curve index for the X-ray afterglows
at 3-time epochs after the GRB.  The solid curves show the expected ``closure relations''
for an adiabatic expanding external shock emitting synchrotron radiation \citep{spn98,chevNli00,sph99}.
Only after $t=2 \times 10^4$s --- a time corresponding to all pre-{\it Swift}~X-ray afterglow
observations --- are the data overwhelmingly consistent with the models.}
\label{fig:closure}
\end{figure}

We show in BK07 that powerlaw energy fits are inappropriate at early
time.  This can also be seen in the first panel of Figure \ref{fig:closure}, which plots
the energy index $\beta$ versus the temporal index $\alpha$ 
for 59 {\it Swift}~afterglows prior to and including GRB~070208 (see Table 1).
For $t<10^3$s, there is a very wide 
range in $\beta$,
consistent with the combined range of low-energy and high-energy indices observed 
in {\it BATSE} GRBs \citep{preece00,kaneko06}.  
Nearly half of these points (42\%) are inconsistent with any of the external shock synchrotron
models plotted as diagonal lines.  The fraction is greater than a half (52\%) if we consider
bursts separately, rather than plotting multiple spectra from individual events.
We show in BK07 that the X-ray spectra are well
fit by the same \citet{band93} model which fits the GRB spectra.  The
large scatter in the time indices for $t<10^3$s is due to rapid light
curve decays and flaring.  In Section \ref{sec:endhard} we show that the X-ray
hardness can be used to infer the end of this phase.

From $10^3\lessim t \lessim 10^4$s, the X-ray light curves typically
decay at a much slower rate.  As shown
in Figure \ref{fig:closure} (middle panel), there is apparently little
spectral evolution.  Most of the fits (79\%) and about half of the
total number of bursts (53\%) are consistent with an adiabatic
shock observed above the cooling frequency $\nu_c$.   \citet{willin06} also find
that about half of all bursts are not consistent with this synchrotron model
due to anomalously slow time decays.  The fraction of consistent spectra is larger, because
the events without plateaus are generally brighter at this stage.
The plateau events can be modelled assuming
a smooth re-injection of energy into the external shock at late time \citep{nousek06,zhang06}
or by deceleration of the external shock in a wind density
($n\propto R^{-2}$) external medium \citep{pain07}.

After $10^4$s, the fits exhibit a tight clustering in
both $\alpha$ and $\beta$ and 88\% of the temporal/spectral 
snapshots (or 80\% of bursts) are consistent with the behavior expected from an adiabatically
expanding shock in the circumburst medium, emitting synchrotron radiation above the
cooling frequency with electron index $p\approx 2$ \citep[e.g.,][]{spn98}.  The pre-{\it Swift}~X-ray
data from Beppo-SAX \citep[see, e.g.,][]{pasquale06} and Chandra and
XMM \citep[see, e.g.,][]{gendre06} are all taken beginning after this time
and show closely consistent behavior with the {\it Swift}~events plotted here.
Apparent here, but discussed in detail in \citet{willin06} and \citet{pain07}, 
few of these events here fit the expectation for a jetted afterglow \citep{rhoads99}.

\section{$T\sim 10^3$s and the End of Hardness Variations}
\label{sec:endhard}

\begin{figure}[H]
\centerline{\includegraphics[width=5.0in]{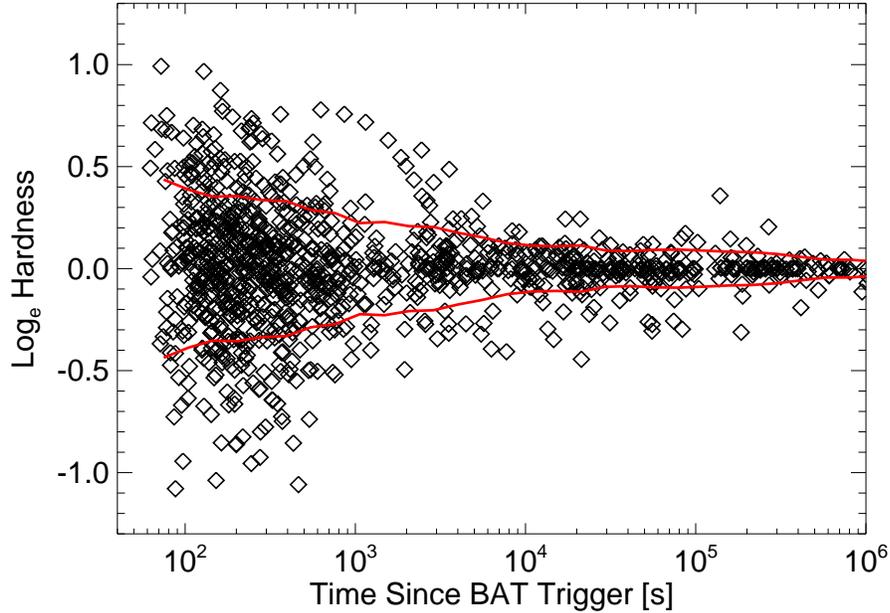}}
\caption{
The scatter in hardness ratio for the full {\it Swift}~sample prior to GRB~070208
drops rapidly with observation time since the GRB trigger.  We have
subtracted off a constant $HR$ from each burst, which we obtain by fitting
the data at $t>10^4$s.}
\label{fig:hr_scatter}
\end{figure}

Figure \ref{fig:hr_scatter} plots the X-ray hardness ratio versus time
since the GRB trigger for the full sample of XRT afterglows.  We also show
$\pm1$ times the root-mean-square scatter in the data in red.  The scatter
is several times greater prior to $10^4$s than after.  The reason for this scatter
becomes clear when looking at the afterglows from individual bursts.
Figure \ref{fig:hr_multi} shows the X-ray light curves and coincident
X-ray hardness ratios for six events.  Each shows an early period of
strong hardness variation, which flattens out
to a late-time value after several hundred seconds.
The hard, late-time component appears to overwhelm the soft, early-time 
component.  We mark as the $T_H$ the time at which the hardness reaches
a minimum in each plot, before gradually increasing to a constant late-time value.

Figure \ref{fig:spec_time} plots the distribution of end
times of these hardness variations $T_H$ for 50 afterglows (also
Table 1).  The
flux prior to $T_H$ is typically far softer than that after the start of the
plateau phase, and the plateau phase becomes evident in $HR$ plots prior to becoming
evident in the 0.3--10.0 keV count rate.   
We note that the 0.3--10.0 keV light curves transition smoothly across $T_H$.  Therefore
the energy integrated light curve cannot be used to measure $T_H$.
Several additional events, with and without flaring light curves,
are plotted in BK07.  

There is a significant anti-correlation between 
the time since the end of the BAT emission $T_H-T_{\rm 90}$
and the X-ray flux at $T_H$ (Kendall's $\tau_K=-0.46$, signif.$=2\times 10^{-6}$),
which reflects the decay of the GRB flux and its intersection at $T_H$ with
a range of possible flux levels for the afterglow plateau phase component.

The end times $T_H$ are on
average $5\pm3$ longer than the prompt $T_{90}$ durations.
The X-ray flux at $T_H$ is on average $10^{1.9\pm0.8}$ times fainter than the
average flux for the GRB measured in BAT.
These quantities are only weakly correlated: $T_{90}$ versus $T_H$ has a
Kendall's $\tau_K=0.20$, signif.$=0.04$ and $F_{X}$ versus $F_{\gamma}$ has a
Kendall's $\tau_K=0.03$, signif.$=0.71$.

We can use the light curve taxonomy developed by other authors to relate our times $T_H$
to the ``canonical'' \citep[e.g.,][]{nousek06} light curve decay phases.
\citet{obrien06} and \citet{willin06} divide the light curves
into prompt and afterglow phases by fitting models to the energy integrated
light curves.  These authors estimate durations which represent the brightest
and most slowly decaying time regions, excluding the rapidly time-decaying
tails of the emission episodes.  Our $T_H$ values are on average ten
times longer than the prompt time $T_P$ in \cite{willin06}, but there
is no significant correlation ($\tau_K=-0.04$, signif.$=0.75$, for $N=35$ bursts).
There is also no significant correlation between $T_H$ and the \citet{willin06}
$T_A$ ($\tau_K=0.1$, signif.$=0.43$, for $N=35$), which approximately measures the
end of the plateau phase and is on average twenty times larger than our $T_H$.

Given $T_H$ as a dividing line between emission with strikingly different
temporal and spectral characteristics --- which we can interpret as a dividing
line in time between GRB and afterglow emission --- it is possible to
separate and compare the fluence from the GRB and afterglow.  
Figure \ref{fig:flux_flux}
plots the fluence prior to $T_H$ versus the fluence after $T_H$.  The
afterglow emits an amount of energy proportional to ($\tau_K=0.35$, 
signif.$=3\times 10^{-4}$) and $10^{1.0\pm0.5}$ times lower than the GRB.
The fluence in the X-ray band prior to $T_H$ contributes only 10\%
additional fluence, on average, to the GRB as observed in BAT.  These quantities
are possibly weakly correlated ($\tau_K=0.18$, signif.$=0.07$).  The
prompt and afterglow fluences we find here are consistent with those
reported by \citet[][Fig 3]{willin06} using two-component model fits to the
light curves, but our fluences correlate with less scatter (Figure \ref{fig:flux_flux}).

\section{Achromaticity of the X-ray Plateau Phase at $10^3\lessim T\lessim 10^4$s}
\label{sec:achromat}

\begin{figure}[H]
\centerline{\includegraphics[width=7.5in]{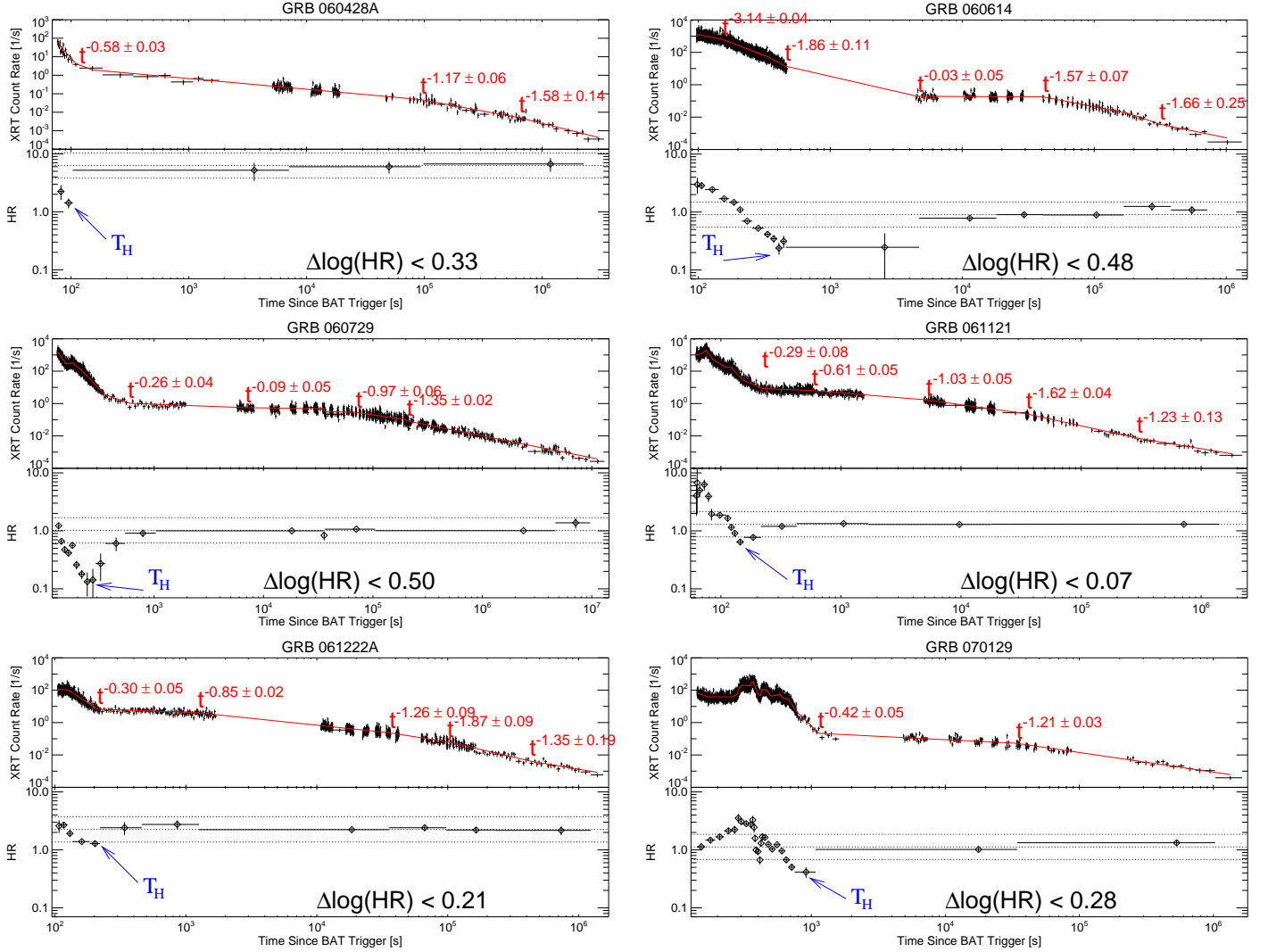}}
\caption{
X-ray Light curve and hardness ratio $HR$ plots for several GRB afterglows 
with prominent light curve ``plateau'' phases.  The temporal indices
of the best fit temporal powerlaw models (red curves) for the counts rates
are shown.  $T_H$ marks the end of an early phase of strong hardness
evolution in each case.  At later time, $HR$ is consistent with constant.
The dotted lines mark the best-fit late-time value as well as 
$e^{0.5}$ and $e^{-0.5}$ times the best-fit value, as a characteristic
expected range for variations in external shock synchrotron models.
A limit on the maximum deviation (1-sigma) from the best-fit late-time
$HR$ value is also given.}
\label{fig:hr_multi}
\end{figure}

\begin{figure}[H]
\centerline{\includegraphics[width=4.0in]{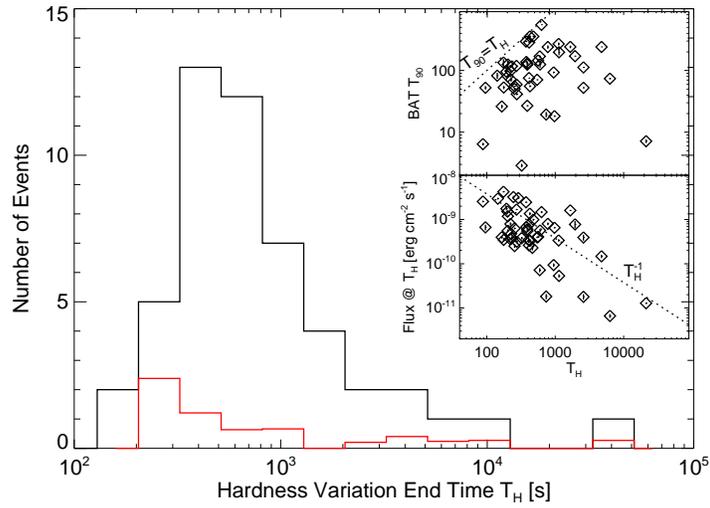}}
\caption{
Frequency distribution (black histogram) of the latest time at which
spectral hardness
variations are observed $T_H$ (Table 1) for 50 events.  The hardness typically
remains constant after this time (Figure \ref{fig:hr_multi}).  The red histogram for fewer
events is divided through by the event redshift ($1+z$).  The subpanel shows the
$T_{90}$ duration measured in BAT versus the hardness variation end-time and also
the X-ray flux at $T_H$ versus $T_H$.  There is a weak correlation with $T_{90}$ and
a strong correlation with $F_X$.  On
average, $T_H$  is 5 times greater than $T_{90}$ and the X-ray flux at $T_H$ is
175 times below the average GRB flux observed by BAT.}
\label{fig:spec_time}
\end{figure}

\begin{figure}[H]
\centerline{\includegraphics[width=4.0in]{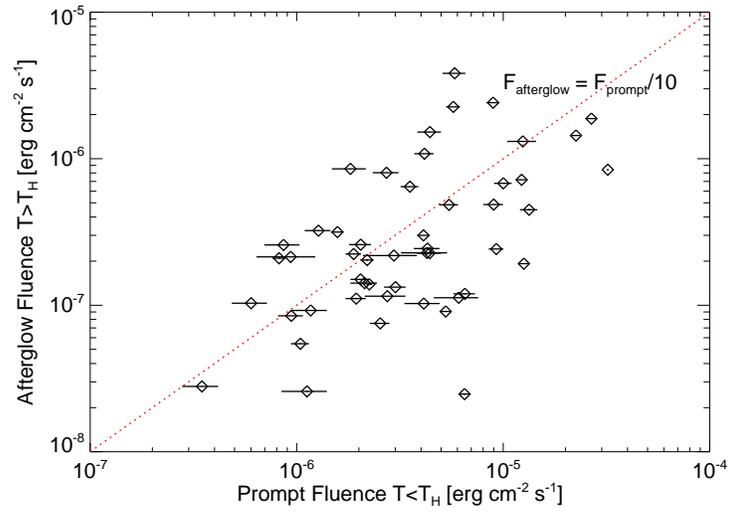}}
\caption{
Time integrated flux before $T_H$ compared to after $T_H$.  The fluence
in the afterglows ($T>T_H$) component is on average ten times less than
the fluence in the prompt GRB and early X-ray emission.}
\label{fig:flux_flux}
\end{figure}

The events selected for plotting in Figure \ref{fig:hr_multi} have prominent
flat X-ray light curves at $10^3\lessim t\lessim 10^4$s.  
On the $HR$ panel for each burst in Figure \ref{fig:hr_multi} we print the
maximal $HR$ variation between data points after $10^3$s.  This is always less than
0.5, corresponding to $\delta \beta < 0.5$.  $\delta \beta = 0.5$ is the expectation for
changes due to cooling in the external shock \citep[e.g.,][]{spn98,chevNli00}.  If we
place a tighter constraint on the allowed hardness ratio variations by allowing only
powerlaw increases or decreases after $10^3$s, then the limits on $\delta \beta$ are
much tighter, $\delta \beta \lessim 0.1$.  We thus see no evidence for significant spectral
evolution in these events after $10^3$s.

We also observe no significant variation in $\beta$ from powerlaw fits, consistent with the $HR$
analysis.  Combining this information to derive a constant $\beta$ in time for each event,
we can plot $\alpha$ and $\beta$ across the break (Figure \ref{fig:plateau_indices}).  Many of the
light curve breaks in Figure \ref{fig:hr_multi} are gradual and are fit here with multiple
powerlaw segments.  As seen in Figure \ref{fig:plateau_indices} and discussed more below, the 
values of $(\alpha, \beta)$ during the plateau are consistent with values produced in energy 
injection models.  As the light curve breaks, the $(\alpha, \beta)$ approach those expected 
from external shock models without energy injection.

\section{Discussion}
\label{sec:discussion}

We have exploited an autonomous spectral/temporal division of early afterglow data
to isolate the time when models for the GRB afterglows well-tested prior to the launch of 
{\it Swift}~first begin to match the data well.   From powerlaw fits in time and energy (Section \ref{sec:temp_spec}), the afterglow
models appear to break down strongly prior to $t\lessim 10^4$s.  This is consistent with
the findings of \citet{obrien06,willin06}.  We focus here on rapid time variations in the
X-ray hardness, which end by $T_H\approx 10^2-10^4$ s and therefore allow for a clean separation
of early, GRB-like emission and later afterglow-like emission without hardness variations.

Our finding here that the end time of hardness variations $T_H$ anti-correlates strongly with the X-ray flux at
$T_H$ (Figure \ref{fig:spec_time}) likely has a trivial explanation: because the light curve is decreasing logarithmically after $T_{90}$,
the length of the duration $T_H-T_{90}$ simply reflects the faintness of the afterglow.  
More interesting,
the large dynamic range in this correlation between prompt and afterglow fluences (Figure \ref{fig:flux_flux}) indicates that some physical feature of the explosion or circumburst
cite must be able to substantially modulate the fraction of energy in highly relativistic material (the GRB)
or the shock kinetic energy (the X-ray afterglow).

GRB models must be able to explain
how the fireball deceleration \citep[see,][and references therein]{mes02} can be postponed until after $t\approx 10^3$ s --- and probably until after $t\approx 10^4$ s --- when the
afterglow light curve is no longer flat or rising.  There must be no apparent imprint of the external medium on the light curve or
spectrum prior to $t\approx 10^3$ s.
Because the deceleration time
$t_{\rm dec} =390(1+z)E_{{\rm shock},53}^{1/3}n_1^{-1/3}\Gamma_{2}^{-8/3}$ s \citep[e.g.,][]{piran99,rmr01},
this might be accomplished by increasing in the shock kinetic energy $E_{{\rm shock},53}$ [$10^{53}$ erg] via energy injection, by having a very low density $n$ [cm$^{-3}$],
or by placing much of the outflow in low Lorentz factor $\Gamma = 100 ~\Gamma_2$ material.
The importance of these parameters becomes more apparent if we focus on extreme cases (Sections \ref{sec:prior} and \ref{sec:shock}) or on modelling of the
more typical cases (Section \ref{sec:inject}).

\subsection{Events with No Hardness Variations}
\label{sec:prior}

There was an expectation, based primarily on detections of putative afterglow components
in the tails of GRBs \citep{con02,giblin02,lrg01} and also from studies extrapolating the X-ray afterglow
flux back to the prompt emission \citep[e.g.,][]{costa97,piro98,frontera2000}, that the X-ray afterglows
measured by \swift~would directly proceed and connect to the prompt emission.  That is, the emission
at $t\approx 10^2-10^4$s was expected to mirror the temporal/spectral properties of the well-established later
emission plotted in the third panel of Figure \ref{fig:closure}.
Instead, the rapid time-variation and plateau phases generally occur, with strong spectral variations
during the rapid time-variation phase.  It is a well known fact that the late-time emission in {\it Swift}~GRB
X-ray afterglows does not typically extrapolate back in time to the end of the prompt emission \citep[e.g.,][]{nousek06,zhang06,obrien06,willin06}.
Instead, the X-ray light curve is relatively flat prior to $t\approx 10^4$s, and this results in typical X-ray fluences
that are ten times lower than the GRB fluence \citep[Figure \ref{fig:flux_flux}, also,][]{zhang06b,gkp06}.
Because the late time X-ray emission is typically argued to trace the kinetic energy explosion \citep{k2000,fnw01},
this may lead to an efficiency problem because the shock energy available to the GRB is very low \citep[e.g.,][]{piran99}.

Because efficiency of the conversion of shock kinetic energy to the energy released by the GRB is an intrinic property,
we would expect that the efficiency should vary little from event to event and that all events should exhibit an X-ray
plateau.  However, there are few bursts --- $<$10\%
of the 30 or so afterglow observations which began early and measured with high signal-to-noise
the tail of the prompt emission --- which do not show early $HR$ variation and do appear to show an
afterglow-like component decaying as a powerlaw after the prompt emission at $t\gtrsim 100$s.  The  best
example GRBs 050717 \citep[also,][]{krimm06}, 060105 \citep[also,][]{tash06}, and 
061007 \citep[also,][]{mundell07,schady07} are shown in Figure \ref{fig:hr_multi2}.
These have especially bright and hard prompt emission with $E_{\rm peak}\gtrsim 500$ keV
(suggestive of high $\Gamma$)
and hard X-ray emission detected beginning after $t\approx 90$s with energy index $\beta \approx 1$.  
Perhaps an early deceleration occurs for these events due to an anomalously high circumburst density.
Could these events also be telling us that energy is present but not observed in the soft X-ray band in
the other, more common afterglows with prominent plateau phases?  Where is this energy?

One possibility is the two jet model \citep{eg06}, with a GRB jet containing more
kinetic energy per solid angle than in the afterglow.  \citet{pain07} has tested and found no evidence for
this scenario.  Another possibility is that the early forward shock emission is suppressed as it scatters
(to the GeV-TeV range) photons from the late-time internal shocks and flares \citep{wlm06}.
Understanding a possible energy removal mechanism and its impact on the fluence correlation (Figure \ref{fig:flux_flux}) will likely help also to understand a possible
correlation between the end of the plateau phase and the GRB energy reported in \citet{obrien06} and
\cite{nava07}.

Finally, we note that GRBs 050717, 060105, and 061007 may be representative of a separate class of GRBs to which previous missions were more sensitive.   
This would explain why afterglow-like tails are rarely observed just following {\it Swift}~GRBs.
In part, previous mission were likely also less sensitive to the very soft emission detected by {\it Swift}.  A handful of these ---
050502B, 050724, 061222B, 070129 --- have very soft emission between $T_{90}$ and $T_H$ which is greater than the prompt fluence.
These events may help us to understand how X-ray Flashes \citep{heise2000} are related to GRBs.  A late and bright X-ray flare as in 050502B
likely produced the soft X-ray excess in the enigmatic GRB~031203 \citep{watson06}.

\subsection{Events with Very Late-time Hardness Variations, The External Shock?}
\label{sec:shock}

There are rare events which show hardness variations with $\Delta HR>0.5$ after $t\approx 10^3$s (Figure \ref{fig:late_hard2}),
which is probably too late for an explanation involving the deceleration of the GRB fireball.
The unusual supernova-GRB~060218 stops varying in hardness just after $10^4$s.
(Given the extremely long prompt duration of this event
\citep[$T_{90}\approx 2\times 10^3$s;][]{taka06}, the late hardness evolution may not be unusual.
This GRB and afterglow produces a clear arc from $t\approx 300-3000$s in
Figure \ref{fig:hr_scatter}, which demonstrates a $HR$ values and evolution distinct from those observed in any other event.)
The high-$z$ GRB~050904 light curve appears to consist entirely of flares,
and this is reflected in late-time hardness variations.

The hardness increases at late-time for GRB~060206 (the outlier in Figure \ref{fig:spec_time}).
The additional cases (GRBs~050315, 060105, 060814) show a decreasing hardness on a timescale similar to the observations time.  The hardness
variation is consistent with the factor of two expected from the $\Delta \beta=0.5$ change
expected from a cooling break in the synchrotron shock picture \citep[e.g.,][]{spn98,chevNli00}.
Although we believe we have accounted for the flux of a nearby source, it is possible that the hardness
increase in the case of GRB~060206 is due to contamination by that source.  This is not an issue for GRBs 050315, 060105, and 060814.

For GRB~060206 at $z=4.045$ \citep{fynbo06}, the source-frame GRB energy release is
$E_{\rm iso} = 4.2^{+0.8}_{-0.6} \times 10^{52}$ erg.
The break to increased hardness in Figure \ref{fig:late_hard2} for a wind medium implies a reasonable density $A_*  = 0.13 (\epsilon_B/0.01)^{-3/4}$.
If the density were uniform, as would be inferred from the breaks to softer spectra in the other 3 events,
a very low density $n= 1.7 \times 10^{-5} (\epsilon_B/0.01)^{-3/2}$ is implied.  Unless the other 3 events are
at low redshift ($z\lessim 0.1$), which is unlikely given the lack of bright optical emission in each case,
the implied densities are anomalously low.

Falling back on our basic ignorance of the nature of the prompt engine and its timescales for energy output, the late time
variation could also be due to the central engine.   The light curves do appear more structured than simple powerlaws.
In Figure \ref{fig:late_hard2}, we fit the observed 0.3--10 keV rate model to the $HR$ in order
to derive a hardness intensity correlation index.  These are printed in the figure and are typically
$\approx 0.5$ for the early, GRB-like emission and $\approx 0$ for the late emission.  As we discuss in
BK07, a probable explanation for this correlation at early times is a relativistic viewing effect
due to photon arrival from emitting regions off the line of site.
The long emission timescales could be due to curvature radiation at large radii.  It is interesting to note
that the late hardness evolution in GRB~050315 rules out associating the late break with an achromatic jet break, as was done in
\citet{vaughan06}.

\subsection{Energy Injection Scenarios}
\label{sec:inject}

The lack of spectral evolution during the light curve plateau phase and at later time suggests strongly 
that these episodes are generated by the same emission mechanism.
In the fireball model, this points to a smooth late-time energy
injection that refreshes the external shock \citep{nousek06,zhang06}, which alters only the time decay
rate and not the spectral regime in which the X-ray synchrotron emission occurs.
The external shock without energy injection cannot produce the observed flat light curves \citep[e.g.,][]{spn98}.

\begin{figure}[H]
\centerline{\rotatebox{270}{\includegraphics[width=3.5in]{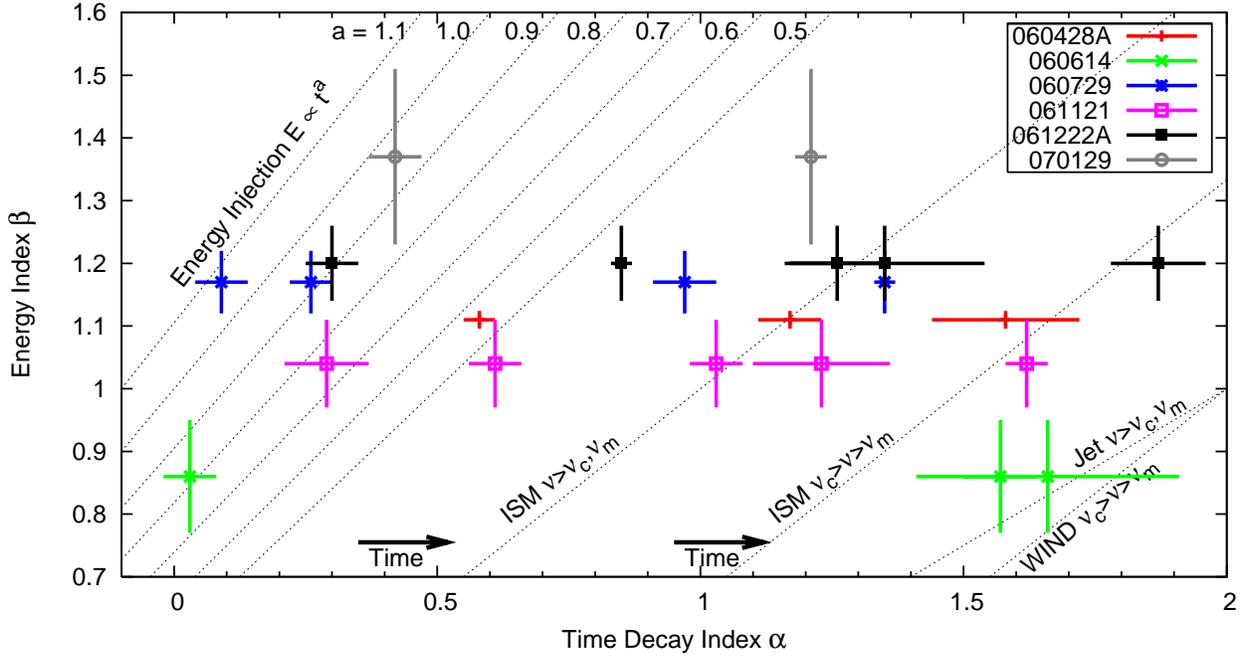}}}
\caption{
Energy and time indices for powerlaw segments in the six bright plateau
events from Figure \ref{fig:hr_multi}.  Separate events are color coded.
Indices for the flat, plateau phase of the light curve in each case
are consistent with energy injection models (dotted lines shown are for $\nu>\nu_c,\nu_m$).  
The break in time to a steeper fade --- which occurs roughly after the
model marked $a=0.5$ in this plot --- is gradual.  The powerlaw indices
at late time approach those expected from the external shock models.  Models
for the late time emission of spherical and jetted shocks are plotted.}
\label{fig:plateau_indices}
\end{figure}

\begin{figure}[H]
\includegraphics[width=6.8in]{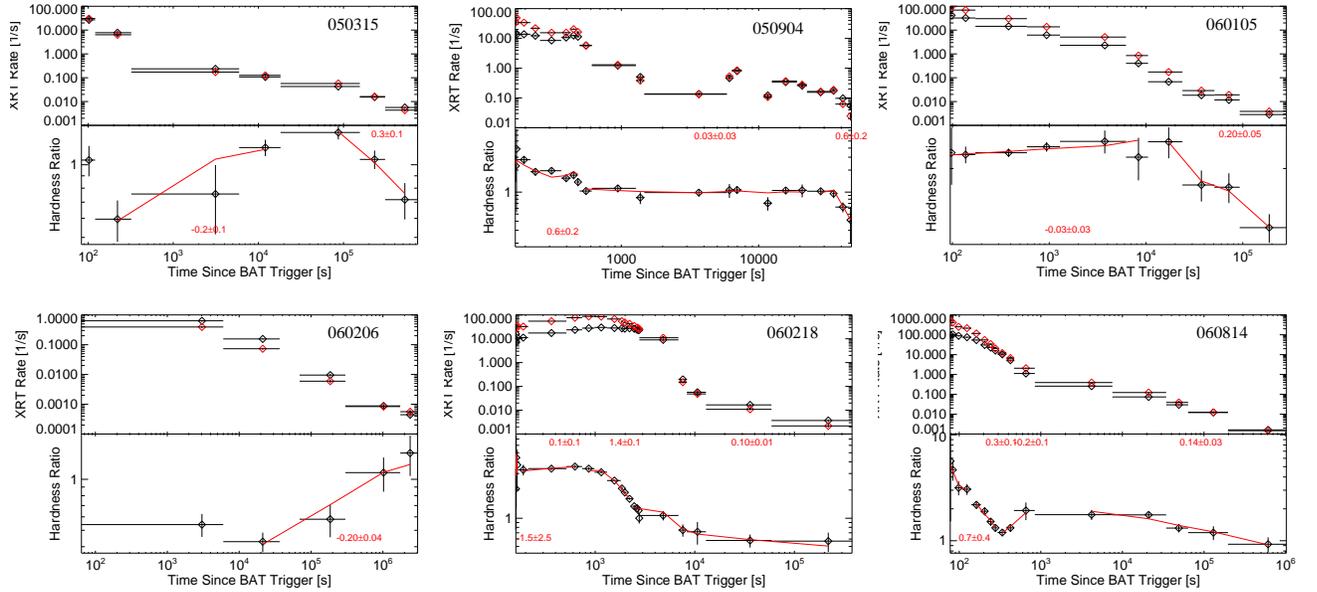}
\caption{
X-ray Light curve and hardness ratio $HR$ plots for rare GRB afterglows
which show late-time hardness variations. 
These variations may imply very late time central engine activity.
Fits of the observed flux to $HR$ are plotted in red, and index values
$\approx 0.5$ are expected for internal shocks \citep{bNk07}.}
\label{fig:late_hard2}
\end{figure}

\begin{figure}[H]
\centerline{\includegraphics[width=4.0in]{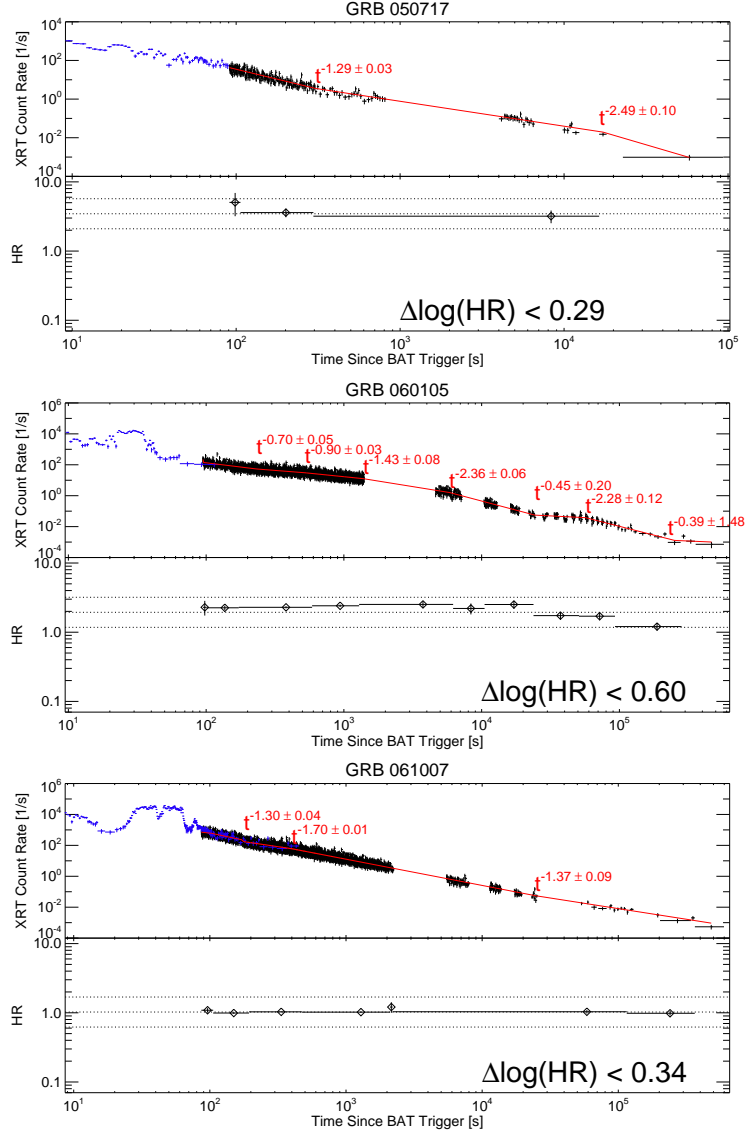}}
\caption{
X-ray Light curve and hardness ratio $HR$ plots for three events
where the GRB tail (blue points) connects directly with a powerlaw (afterglow-like) X-ray light
curve exhibiting little spectral variation.  The X-ray data in these cases are
well fit by powerlaws with energy index $\beta=1.0\pm0.1$ throughout.
The dotted lines mark an
expected range for variations in external shock synchrotron models (also Figure \ref{fig:hr_multi}).
In the case of GRB~060105, the 1-sigma limit on $\Delta \log(HR) \propto \Delta \beta/0.9$ is consistent with a possible
cooling break at $t\approx 10^4$s.}
\label{fig:hr_multi2}
\end{figure}

Energy injection into the GRB external shock was first discussed by \citet{p98} and \citet{rm98}.
According to the very general energy injection models outlined in \citet{nousek06}, the X-ray light curve
is propped up by an insertion of energy $E$ in time as $E\propto t^a$.  For data observed above the
synchrotron cooling break $\nu_c$ and peak frequency $\nu_m$, the flux varies as $F_{\rm X} \propto 
t^{-\alpha_{\rm inj}}$, with $\alpha_{\rm inj} = \alpha - a(\beta+1)/2 = (3-a)\beta/2 - (1+a)/2$.
Curves for $a$ in the range $0.5-1.1$ --- which is needed to account for the well-sampled plateau phase events 
described in Section \ref{sec:achromat} --- are shown in Figure \ref{fig:plateau_indices}.  
Time flows from the left to the right in this figure as the light curve smoothly breaks.  We consider the left-most
points for each burst as those most likely to reflect the true energy injection profile.
If the X-ray band is below the cooling break, the slopes in Figure \ref{fig:plateau_indices} remain the same for the 
same $a$ but the offsets shift. This leads to acceptable fits with $a=0.7-1.1$ for a constant density (ISM) medium 
and $a=1.0-1.4$ for a wind density medium.  An $E \propto t$ scaling implies a central
engine with an approximately constant late-time luminosity or an ordered flow of internal shock
material with $M(>\Gamma) \propto \Gamma^{-2}-\Gamma^{-5}$ (ISM medium) reaching the afterglow shock at late time \citep{nousek06,gkp06}.

After the break, the fits become consistent with external
shock models ranging between the expectation for a spherical expansion observed at $\nu>\nu_c,\nu_m$ and
a jetted expansion in the same regime.  This may imply that gradual jet breaks are present in these events,
although only one (060614) ever reaches the expected late-time decay rate.

Alternatively, the X-ray band could be in the $\nu_c>\nu>\nu_m$ regime and the energy injection
could be turning off gradually.  The steep decay in the case of 060614 at late time requires a wind
density medium.  However, this can be ruled out from optical data reported to the 
GCN\footnote{http://gcn.gsfc.nasa.gov}. During the plateau,
the optical light curve rise as $t^{0.38\pm0.06}$, as compared to $t^{-0.03\pm0.05}$ measured in the
X-ray band.  This behavior is consistent with $\nu_c$ between the optical and X-ray band and energy
injection with $a=0.69\pm0.08$.  There are no optical points after the candidate jet break to verify achromaticity.
GRB~060729 has consistent optical ($t^{-0.24\pm0.03}$) and X-ray ($t^{-0.26\pm0.04}$) light
curve indices during the plateau, and both light curves break to consistent decays thereafter.
The post break decay is consistent with expansion into an ISM medium without a jet break.
The energy injection prior to the break is fit by $a=0.92\pm0.05$ ($\nu>\nu_c,\nu_m$).

\citet{pain07} propose a very simple model to explain some plateau light curves and spectra.  
For a wind density medium, $t_{\rm dec}$ is a strong function of the bulk Lorenz factor, $t_{\rm dec}
=6(1+z)E_{{\rm shock},53}A_*^{-1}\Gamma_{2}^{-4}$ s for
a typical Wolf-Rayet wind density of $5 \times 10^{11} A_*$ g cm$^{-1}$.
The afterglow will not peak until $10^3-10^4$s if, after the internal shocks are through,
the effective $\Gamma\sim 20$.  During deceleration, the flow coasts and the light curve
stays relatively flat, $\alpha=\beta-1$ \citep{pain07}, for $\nu>\nu_c,\nu_m$.  Contrarily,
deceleration by a uniform density medium produces a sharply rising light curve, which is not observed.
This $\alpha,\beta$ relation for the wind medium takes the
same form as that for the $a=1$ energy injection model, and it appears to be roughly consistent 
with most of the plateaus in Figure \ref{fig:plateau_indices}.  In this picture, the jet break
will coincide with the end of the plateau if the opening angle is $\theta \approx 1/
\Gamma \approx 3$\degree, which may well occur for some events.

Several additional models have been proposed to explain the X-ray plateau phase, and many of these can be constrained
by a constant X-ray spectral slope and from the fact that a distinct hardness evolution separates the plateau phase
emission from the GRB-like emission prior.
One possibility --- which we can rule from lack of hardness variations because it requires contribution from the
spectrally varying tail of the GRB --- is that off-axis afterglow plus late GRB emission combine to produce the plateau
\citep{eg06}.  In a similar fashion, we may be able to rule out the ``late prompt'' model of \citet{gglf07}, although
spectral variations would be modest in that model due to low $\Gamma$.
Models involving the reverse shock \citep{gdm07,ub07} appear to produce spectra variations in the X-ray band.
Inverse Compton models which extract afterglow flux should also change the spectrum, but see \citet{wlm06}.
A final, more exotic possibility which we cannot rule out is time
evolution of the microphysical parameters defining the shock \citep{gkp06}.

\section{Conclusions}
\label{sec:concl}

We have shown that GRB and early afterglow light curve prior to $T_H\sim 10^3$s is highly time and energy variable.
The flux at the end of these variations and during the X-ray plateau phase exhibits non-changing X-ray
hardness like the latest X-ray afterglow emission that is well modelled by a synchrotron external shock
($t\gtrsim 10^4$s; Section \ref{sec:temp_spec}).
The afterglow flux is typically ten times lower than would be estimated from a simple
extrapolation of the GRB flux after $T_{90} \sim T_H/5$.  Explaining how the GRB deceleration can be postponed until after 
$t\gtrsim 10^4$s is a central challenge to those modelling GRB and their afterglows.

Observations prior to \swift~which imply a common early onset of the afterglow may point to a
class of bursts rarely observed by \swift.  The three \swift~examples discussed (GRBs 050717, 060105, and 061007) 
have energetic and hard prompt emission.  These bursts may be those most rapidly decelerated by the circumburst medium.
These events likely have the most high energy photons for {\it GLAST} to observe, unless photons missing from the
early afterglows of softer events are preferentially up-scattered to high energies by late time shocking \citep[see, e.g.,][]{wlm06}.
In either cases, {\it GLAST}~observations will be crucial for understanding this diversity and for helping us to understand
the {\it Swift}~phenomenology relative to that observed in previous missions \citep[see, e.g.,][]{zhang07}.
Additional long wavelength observations (e.g., in the optical/IR) are also essential at times $t\lessim 10^3$s,
because these better probe the circumburst density structure.

\acknowledgments
N.~R.~B gratefully acknowledges support from a Townes Fellowship at U.~C. Berkeley Space Sciences Laboratory and partial support
from J. Bloom and A. Filippenko.  
D.~K. acknowledges financial support through the NSF Astronomy $\&$ Astrophysics Postdoctoral Fellowships under award AST-0502502.
Special thanks to the {\it Swift}~team for impressively rapid public release and analysis of the XRT data.  
Thanks to J. Bloom and the U.~C. Berkeley GRB team for comments on the manuscript and several useful
conversations.  We thank an anonymous referee for a very useful and critical reading of the manuscript.

\clearpage

\begin{table}[H]
\begin{center}
\caption{X-ray Hardness Evolution End Times $T_H$}
\vspace{5mm}
\footnotesize
\begin{tabular}{lclclclc}\hline\hline
 GRB & $T_H$ [s] & GRB & $T_H$ [s] & GRB & $T_H$ [s] & GRB & $T_H$ [s] \\\hline
050219A & $165^{+35}_{-35}$ & 050315 & $221^{+100}_{-100}$ & 050502B & $974^{+82}_{-82}$ & 050607 & $410\pm40$ \\
050712 & $428^{+59}_{-59}$ & 050713A & $203^{+31}_{-31}$ & 050714B & $400^{+4300}_{-20}$ & 050716 & $544^{+170}_{-170}$ \\
050721 & $1135^{+242}_{-242}$ & 050724 & $323^{+38}_{-38}$ & 050730 & $779^{+11}_{-11}$ & 050801 & $737^{+142}_{-142}$ \\
050814 & $562^{+275}_{-274}$ & 050820A & $4750^{+100}_{-100}$ & 050822 & $600^{+32}_{-32}$ & 050922B & $1121^{+129}_{-129}$ \\
051117A & $1659^{+54}_{-54}$ & 051227 & $220\pm100$ & 060115 & $594^{+152}_{-152}$ & 060204B & $378^{+37}_{-37}$ \\
060206 & $21284^{+15275}_{-15275}$ & 060210 & $426^{+7}_{-7}$ & 060211A & $391^{+88}_{-87}$ & 060312 & $176^{+29}_{-29}$ \\
060418 & $175^{+12}_{-12}$ & 060427 & $226^{+26}_{-26}$ & 060428A & $95^{+8}_{-8}$ & 060428B & $271^{+35}_{-35}$ \\
060510B & $415^{+21}_{-21}$ & 060526 & $374^{+16}_{-16}$ & 060604 & $193^{+4}_{-4}$ & 060607A & $289^{+7}_{-7}$ \\
060614 & $2588^{+2127}_{-2127}$ & 060707 & $6300^{+565}_{-565}$ & 060708 & $88^{+11}_{-11}$ & 060714 & $232^{+12}_{-12}$ \\
060719 & $255^{+147}_{-147}$ & 060729 & $271^{+26}_{-26}$ & 060904A & $413^{+47}_{-47}$ & 060904B & $1948^{+1652}_{-1652}$ \\
060929 & $630^{+11}_{-11}$ & 061110A & $390^{+26}_{-26}$ & 061121 & $142^{+11}_{-11}$ & 061202 & $390^{+199}_{-199}$ \\
061222A & $203^{+18}_{-18}$ & 061222B & $275^{+48}_{-48}$ & 070107 & $477^{+29}_{-29}$ & 070110 & $254^{+100}_{-101}$ \\
070129 & $948^{+171}_{-171}$ & 070208 & $2585^{+2175}_{-2175}$ & & & & \\\hline
 ~~Limits & & & & & & & \\
 050401 & $<$153 & 050416A & $<$126 & 051016B & $<$91 & 051109A & $<$195 \\
 060111B & $<$145 & 060211B & $<$95 & 060219 & $<$158 & 060502A & $<$88 \\
 060510A & $<$110 & 060906 & $<$276 & 060908 & $<$90 & & \\\hline\hline
\end{tabular}
\end{center}
{\footnotesize Notes: Several afterglows allow for only a $T_H$ limit measurement.}
\label{tab:htimes}
\end{table}

\clearpage

\end{document}